\documentstyle[prb,aps,epsf,twocolumn]{revtex} 
\begin{document}

\title{Magnetic frustration in a  stoichiometric spin-chain compound, 
Ca$_3$CoIrO$_6$}

\author{S. Rayaprol, Kausik Sengupta  and  E.V. Sampathkumaran \footnote{Corresponding author: sampath$@$tifr.res.in}}
 
\address{Tata Institute of Fundamental Research, Homi Bhabha Road, 
Mumbai - 400 005, INDIA.}

\maketitle

\begin{abstract} 
{The temperature  dependent ac and dc magnetization  and heat capacity 
data of   Ca$_3$CoIrO$_6$, a spin-chain compound crystallizing in 
a K$_4$CdCl$_6$-derived rhombohedral structure, show  the features due to 
magnetic ordering of a frustrated-type below about 30 K, however without exhibiting the signatures of the so-called "partially disordered antiferromagnetic structure" encountered in the isostructural compounds, Ca$_3$Co$_2$O$_6$ and Ca$_3$CoRhO$_6$. This class of compounds  thus provides a variety  
for probing the consequences of magnetic frustration due to topological reasons in stoichiometric spin-chain materials, presumably arising from  subtle differences in the interchain and intrachain magnetic coupling strengths. This compound presents additional interesting situations in the sense that, ac 
susceptibility  exhibits a large frequency dependence in the 
vicinity of 30 K uncharacteristic of conventional spin-glasses, with this 
frustrated magnetic state being  robust to the application of external 
magnetic fields.}
\end{abstract}
$^*$E-mail address: sampath@tifr.res.in
\vskip6mm

\maketitle
Conventionally, the phenomenon of spin glass freezing has been associated with the 
randomness of exchange interactions, mainly due to random distribution of 
magnetic impurities in a non-magnetic matrix and/or due to crystallographic disorder. In 
stoichiometric compounds, one can in principle encounter spin-glass-like magnetic frustration anomalies due to certain topological reasons, for instance, due to antiferromagnetic coupling among the magnetic moments in a triangular lattice. Thus, the investigation of 
glassy effects due to geometric frustration is one of the most interesting 
topics in magnetism and, in this regard, kagome (2D) or pyrochlore (3D) 
lattices remain at the centrestage of this direction of research.\cite{1} 
In this article, we bring out that the spin-chain compounds of the type, (Sr, Ca)$_3$ABO$_6$ 
(A, B= a metallic ion, magnetic or nonmagnetic), crystallizing in the 
K$_4$CdCl$_6$ (rhomhohedral) derived structure (space group R${\bar 3}$c) [See, 
for instance, Refs. 2 and 3] need an attention in this regard. The 
structure is characterized by the presence of chains of A and B ions running 
along c-direction arranged hexagonally {\it forming a triangular lattice}. 
These chains are  separated by Sr (or Ca) ions. Within the chains, 
alternating AO$_6$ trigonal prism and BO$_6$ octahedra share one of the  
faces.  It is generally found that the interchain interaction is 
antiferromagnetic and therefore triangular  arrangement   of magnetic ions 
in the a-b plane may result in magnetic frustration.\cite{4} 
This, coupled with the tunability of   the relative magnitudes of 
interchain and intrachain magnetic coupling strengths as one varies A and/or 
B,  should  provide an opportunity to observe  a variety of  magnetic 
anomalies among this class of compounds.

A survey of the very limited literature suggests that 
this class of compounds is indeed characterized by a host of novel behavior (see, for instance,
Refs. 2 - 14). In particular, the observation of 'CsCoCl$_3$-like\cite{15} partially 
disordered antiferromagnetic (PDA) structure',   in Ca$_3$Co$_2$O$_6$ (Ref. 4) 
and Ca$_3$CoRhO$_6$ (Ref. 9) is quite fascinating; in the PDA structure, 2/3 of the magnetic chains order 
antiferromagnetically with each other and the remaining 1/3 are left 
incoherent in an intermediate T range; however, the magnetism of the latter 
compounds are {\it unique} in the sense that the intra-chain coupling is 
{\it ferromagnetic} unlike in CsCoCl$_3$ in which case this coupling is also 
antiferromagnetic; in addition, the spins in the incoherent  chains are randomly
frozen at a lower temperature, while in the Cs compound a ferrimagnetic 
ordering is observed. In light of these interesting anomalies in Co-based oxides, we considered 
it worthwhile to subject the sister compound, Ca$_3$CoIrO$_6$, which has been previously reported\cite{3} to
undergo antiferromagnetic ordering below 30 K,  for a reinvestigation. 
The present results reveal that there is only one magnetic transition at about 30 K, which is actually of a spin-glass-like frustrated type,  however without  the 
PDA structural characteristics  in the sense described above for    
Ca$_3$CoRhO$_6$ and Ca$_3$Co$_2$O$_6$. 

Polycrystalline Ca$_3$CoIrO$_6$ was prepared by a conventional solid state 
method. The stoichiometric amounts of high purity ($>$ 99.9$\%$)  CaCO$_3$, 
Co$_3$O$_4$ and Ir  were thoroughly mixed in an agate mortar and calcined in 
air at 1173 K for one day. Then the preheated powder was reground, 
pelletized and heated in air for 10 days at 1273 K with intermediate 
grindings as per the procedure in Ref. 3 
and the specimen thus prepared contained an extra line in the x-ray diffraction pattern.
Subsequently, the specimen was subjected to an additional heat treatment at a higher 
temperature of 1423 K for 2 days with an intermediate grinding. The sample 
thus prepared  was found to be single phase and therefore, under our synthetic conditions, 
we require a temperature  higher than that mentioned in Ref. 3 to synthesize the single 
phase material. 
The x-ray diffraction pattern  is found to be in excellent 
agreement with that reported in Ref. 3 establishing ordered arrangement of 
Co and Ir ions. The lattice 
parameters are: $\it a$= 9.160 $\AA$ and $\it c$= 10.890 $\AA$. Dc $\chi$ 
measurements (1.8 - 300 K) in the presence of various magnetic fields and 
isothermal M measurements at several temperatures were performed employing a 
superconducting quantum interference device (Quantum Design). The same 
magnetometer was employed to measure ac $\chi$ (2-200 K) in a frequency 
($\nu$) range 1 - 1000 Hz in an ac field of 1 Oe, in the presence of 
different dc magnetic fields (H= 0, 5 and 40 kOe). The C 
measurements (2 - 60 K) were performed by  semi-adiabatic heat-pulse method 
employing a home-built calorimeter.

\begin{figure}
\centerline{\epsfxsize=6cm{\epsffile{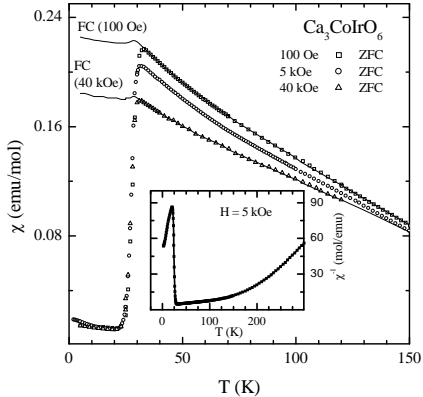}}}
\caption{Dc magnetic susceptibility ($\chi$) as a function of temperature 
(T) for Ca$_3$CoIrO$_6$ measured in the presence of an external magnetic 
field (H) of 100 Oe, 5 kOe and 40 kOe. The data points represent 
zero-field-cooled (ZFC) data, whereas the continuous lines correspond to 
field-cooled (FC) state of the specimen. The FC data for H= 5 kOe  
bifurcates from the ZFC curve at the peak (similar to the curve for other 
two fields) and for the sake of clarity this curve is omitted. Inset shows 
the T-dependence of inverse $\chi$ (ZFC) for H= 5 kOe.}
\end{figure}

The results of dc $\chi$ measurements are shown in Fig. 1 and are found to be 
in broad agreement with those reported in Ref. 3.  $\chi$ 
monotonously increases with decreasing T down to 30 K. The Curie-Weiss law is 
obeyed only above 250 K; there is a continuous change in the slope 
of the plot of inverse $\chi$ versus T (see, Fig. 1 inset) down to about 125 K, below 
which (for T $>$ 30 K) there is  a negligible 
variation in slope. From the high T linear region, the 
effective moment is found to be 4.40 $\mu_B$ per formula unit, which is very 
close to the theoretical value of 4.24 $\mu_B$ under the assumption that Co 
is in 2+ state (S= 3/2, high spin) and Ir is in 4+ state (S= 1/2). The 
paramagnetic Curie temperature ($\theta_p$) is found to be 168 K and the 
positive sign with a large magnitude suggests the existence of strong 
ferromagnetic correlations. As the T is lowered below 30 K, the ZFC $\chi$ 
exhibits a sharp fall, attaining a nearly T-independent value  at lower 
temperatures for all values of H.   The 
field-cooled (FC) $\chi$ curves measured at various fields bifurcate from 
respective ZFC curves at 30 K; this, though sometimes observed  in long range 
magnetically ordered systems, is a characteristic feature of spin glasses.

In order to resolve whether the 30K-feature is due to long range magnetic 
ordering, we have performed ac $\chi$ measurements, the results of which are 
shown in Fig. 2. We first discuss the data taken in zero dc magnetic field. {\it Looking at the curves for
$\nu$= 1 Hz}, there is a 
distinct drop in the real part ($\chi\prime$) of ac $\chi$ as well  below the 
peak temperature of 42 K. There is also a sharp upturn in the imaginary part 
($\chi\prime\prime$) below 45 K  and there is  a distinct peak at 
about 35 K and the plots of ac $\chi$ versus T move to a higher temperature 
with increasing frequency; these 
\begin{figure}
\centerline{\epsfxsize=7.5cm{\epsffile{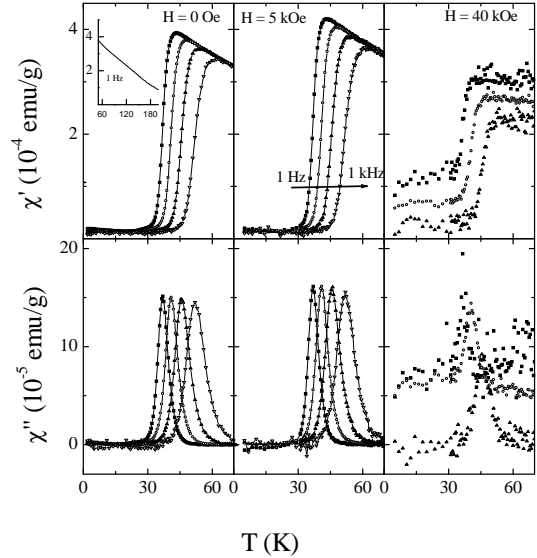}}}
\caption{Ac susceptibility as a function of temperature for various 
frequencies (1, 10, 100, and 1000 Hz) in the absence and in the presence of external dc magnetic 
fields for Ca$_3$CoIrO$_6$. $\chi\prime$ and $\chi\prime\prime$ refer to real and imaginary 
parts. The curves shift to higher temperatures with increasing frequency as marked by an arrow. Gradual and linear variation of the $\chi\prime$ with T in the range 
50 - 250 K is typically shown in the inset ($\nu$= 1 Hz).}
\end{figure}
findings are sufficient enough\cite{16} to 
conclude that there is a spin-glass-like behavior taking place in this T 
range in this compound.  
However, the peak temperature in $\chi\prime$ as 
well as the T-range (32-40 K) where the decreasing part of $\chi\prime$ is sharp 
do not coincide with the corresponding temperatures in dc $\chi$ (30 K and 25-30 K respectively), 
but 
shifted to a higher temperature by about 12 K in $\chi\prime$. 
It is also important to note that the peak temperatures in 
$\chi\prime$ and $\chi\prime\prime$  do not agree. Similar features were observed\cite{17} 
also in pyrochlores with magnetic frustration (due to topological reasons) 
and this implies  strong frequency dependence of susceptibility arising from 
slow magnetic dynamics. Further evidence for this explanation comes from the 
{\it large frequency dependence} of ac $\chi$ peaks; for instance, the curve shifts 
upwards by as much as 15 K for a change of frequency from 1 Hz to 1 kHz. 
Such a large shift corresponds to a value beyond 0.1    for the 
factor,\cite{15} $\Delta$T$_f$/T$_f$$\Delta$(log$\nu$), which is  much 
larger than that known for conventional spin glasses. In this sense the 
dynamics of magnetic frustration is quite unusual. It may be recalled that 
a large frequency dependence of ac $\chi$ has been recently reported 
for Ca$_3$Co$_2$O$_6$ and Ca$_3$CoRhO$_6$ (Refs.  7 and 10). But the magnitude of the discrepancy between the peak temperatures of dc $\chi$(T) and $\chi\prime$(T) is relatively  smaller (about 2 K) for Ca$_3$Co$_2$O$_6$ (Refs. 7 and 14).   

\begin{figure}
\centerline{\epsfxsize=6cm{\epsffile{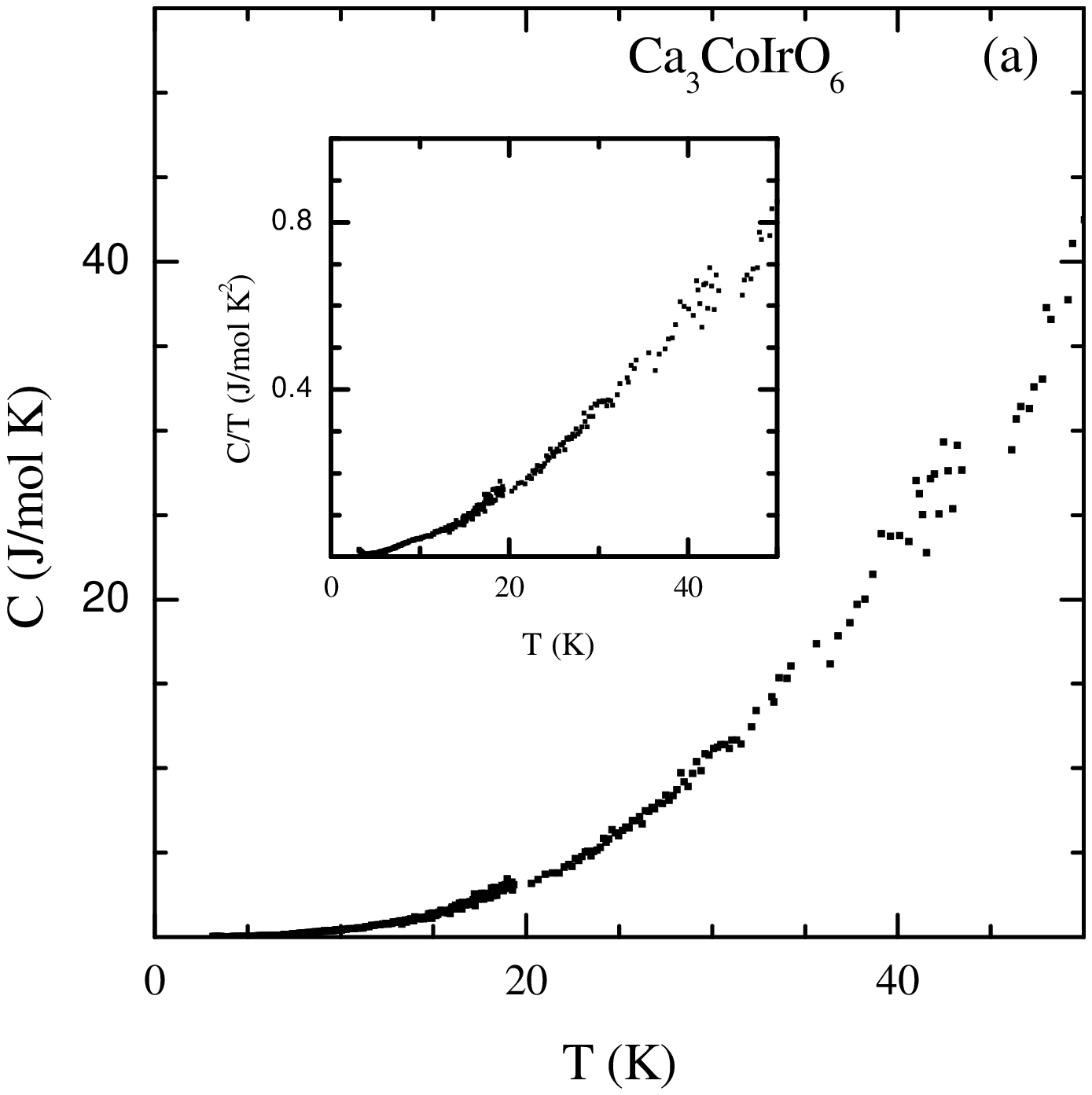}}}
\centerline{\epsfxsize=6cm{\epsffile{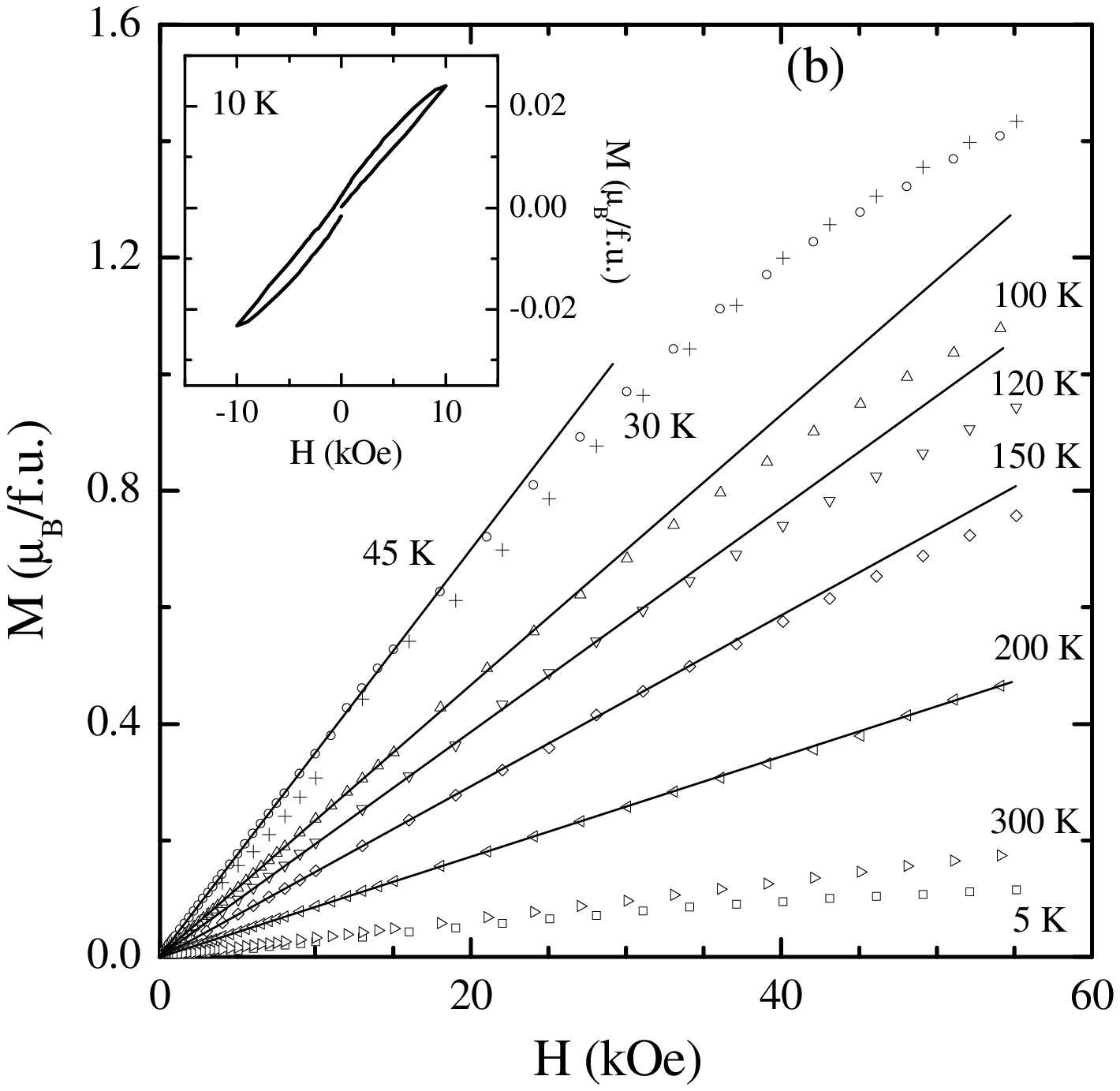}}}
\centerline{\epsfxsize=6cm{\epsffile{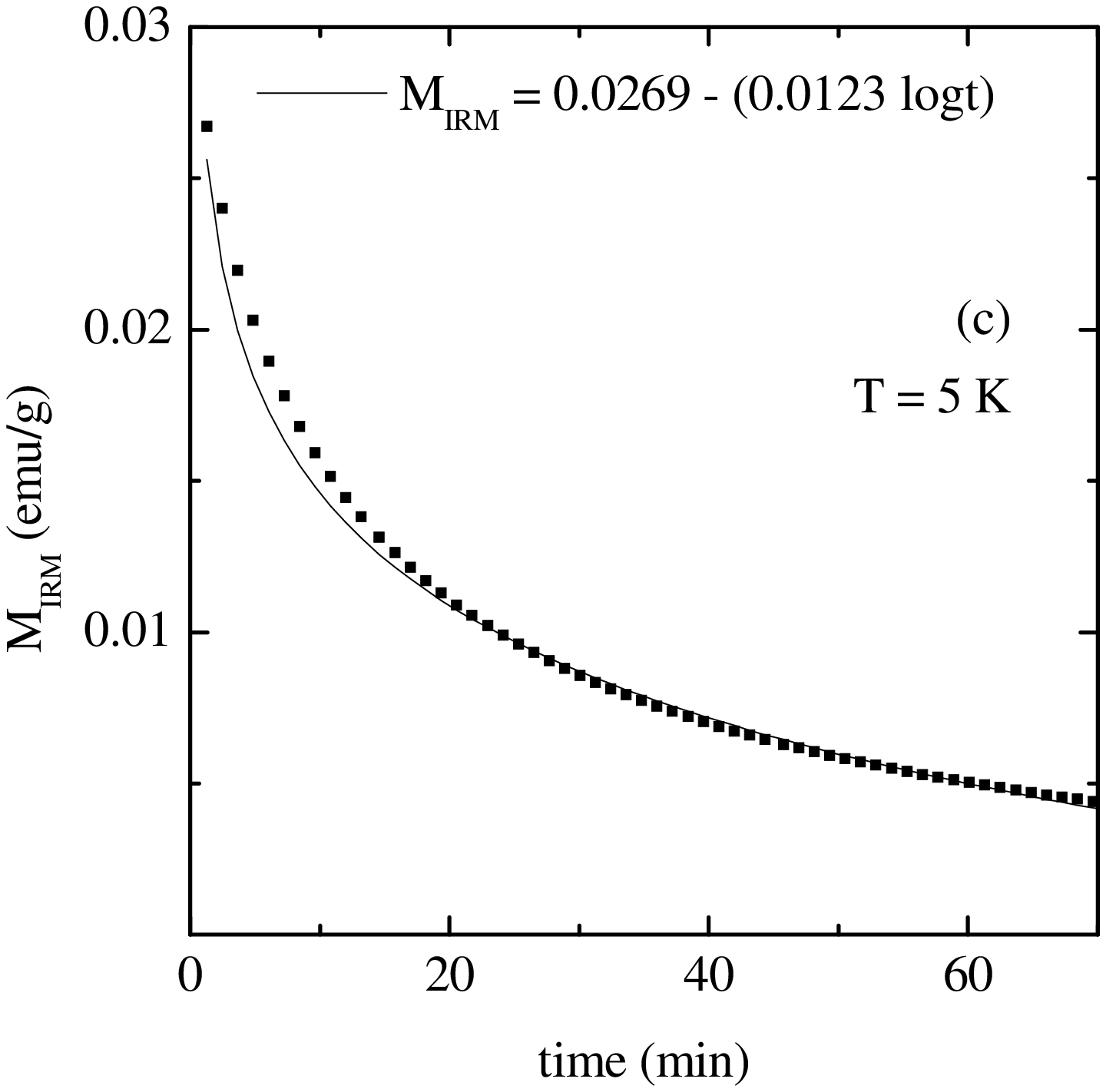}}}
\caption{(a) Heat capacity (C) as a function of temperature (T); inset shows C/T as a function of 
at various temperatures; continuous lines in some cases 
are drawn to highlight low-field linear region; the data for 30 (symbol: +) and 45 K (circles) 
K; and (c) Isothermal remanent magnetization as a function of time (obtained 
as described in the text), for Ca$_3$CoIrO$_6$.}
\end{figure}
Further evidence for spin-glass-like frustration is offered by the C data. It 
is obvious from Fig. 3a that C increases monotonously with T without any 
evidence for a well-defined peak characterizing long range magnetic 
ordering around 30 K. The peak is so washed out that there is no clear-cut 
anomaly even in the plot of C/T (Fig. 3a, inset). It may be recalled\cite{11} 
that an intense peak in C could be seen at 19 K in Sr$_3$ZnIrO$_6$  due to 
antiferromagnetic ordering from Ir ions (S= 1/2), with a magnitude of C jump 
of about 4J/molK.  Therefore,  the absence of a noticeable jump in C around 30 K in 
the present compound conclusively establishes randomness of magnetic 
interactions. Consistent with this, isothermal M below 30 K, say at 10 K 
(see Fig. 3b, inset) is found to be hysteretic with a sluggish 
variation of M with H in the field range of measurement. We have also performed isothermal remanent 
magnetization ($M_{IRM}$) measurements at 5, 45 and 150 K in the following 
way: The sample was cooled from 200 K to the desired T in zero field, then a magnetic 
field of 5 kOe was switched on for 5 mins and the magnetization, $M_{IRM}$, 
was measured as a function of time immediately after the field was switched 
off. While the M attained a very small value (close to the detection limit 
of the magnetometer) at 45 and 150 K, $M_{IRM}$ at 5 K is found to undergo a 
slow relaxation with time, decreasing exponentially for initial periods 
followed by logarithmic variation after waiting for 20 mins (Fig. 3c). This 
slow relaxation of M is also consistent with spin-glass-like freezing.

We have also measured ac $\chi$ in the presence of external dc magnetic 
fields (see Fig. 2). It is fascinating to see that the ac $\chi$ features 
described above, not only   frequency dependence but also the order of 
magnitude of ac $\chi$, are not significantly affected by the field. (The 
plots are found to be noisy for H= 40 kOe and it is so bad  for $\nu =$ 1 kHz 
  that it is not shown in the figure). This situation is completely 
different from that observed for the Rh (for Ir) analogue\cite{10} in the 
sense that the zero-field features are completely suppressed by a field of, 
say, 30 kOe. Thus, the magnetic frustration phenomenon in this compound is 
unique in the sense that it is quite robust to the application of H. This difference 
can be correlated to isothermal M 
behaviour: It may be recalled\cite{10} that isothermal M as a function of H 
in the range 30 - 100 K exhibits a plateau in the Rh compound, whereas, in the 
Ir case, M varies linearly  for initial applications of H followed by a 
curvature towards saturation at higher values of H  without exhibiting a 
plateau (see Fig. 3b). This holds true even in the vicinity of magnetic transition temperature (see the data for 30 K in Fig. 3b).    Thus, it appears that the zero-field magnetic state undergoes 
dramatic modification with H in the Rh case.

A further comparison of M  
behavior of these two compounds in the intermediate T range (30 - 300 K) is 
quite revealing. For instance, for Ca$_3$CoIrO$_6$, in Fig. 1, the 
bifurcation of ZFC-FC $\chi$ sets in at the peak (close to 30 K) 
irrespective of the field, whereas for the Rh analogue, the bifurcation 
occurs at a noticeably higher temperature depending upon the magnitude of 
external H; also, the $\chi\prime$ exhibits a distinct upturn below 100 K for the 
Rh compound as though there is a magnetic phase transition below 100 K, 
whereas $\chi\prime$ surprisingly exhibits a linear decrease with increasing temperature up 
to the highest T measured (see, Fig. 2, inset, for a typical variation). There is a slow decay of M$_{IRM}$, say 
at 62 K as at 5 K in the Rh compound, whereas, in the present case, no magnetic 
relaxation could be observed above 30 K. However, there is a clear evidence for magnetic relaxation at 5 K in the present case as well, as described earlier (see above); it is to be noted that the coefficient of the logarithmic term is larger (about 5 times) compared to that in Ca$_3$CoRhO$_6$, which could mean stronger glassy-effects in the former case. Though dc $\chi$ is field dependent above 30 K
for both the compounds,  this sets in  at a 
{\it well-defined} temperature (close to 100 K, corresponding to the onset of 
PDA structure) for the Rh compound, whereas this 
field-dependence vanishes in a sluggish way persisting  over a wide 
temperature range for the Ir compound. Apparently, the T-range over which this field-dependence  persists may be sample-dependent, as in our case this tendency persists till about 150 K, whereas in Fig. 6 in Ref. 3, this feature appears to terminate around 70 K. Though exact origin of this sample-dependence is not clear, we believe that the T-range over which ferromagnetic correlations (described below) persist may sensitively depend on sample preparative conditions.   

From the above discussion, one can infer that presumably short range 
ferromagnetic correlations persist, say, till about 150 K in the Ir compound without any evidence for glassy features 
in the intermediate T range, 
whereas in the Rh analogue, the signatures of the existence of 'first' long range magnetic ordering (arising 
from intrachain interaction) and glassy behaviour set in at a well-defined temperature near 100 
K.  Interestingly, the plots of inverse $\chi$ versus T above 30 K for both 
look similar in the sense that there is not much change in the slope below 
about 150 K for both. Also, the sign (positive) and the magnitude 
of  $\theta_p$ (close to 170 K) are practically the same for both, as though both should have
similar magnetic characteristics. It is therefore interesting that the present compound does not 
apparently show PDA characteristics as in the case of the Rh compound. The fact that there is
no well-defined C jump at 30 K in the Ir case implies that the even the 30K-transition is {\it purely}
glassy-like in its character, rather than originating from 
simultaneous long range ordering from 2/3 of the chains. It looks as though this difference between these two compounds 
lies in the elongation of magnetic chain of the Ir 
compound ($c$= 10.890$\AA$) with respect to the Rh compound ($c$= 
10.730$\AA$) as the basal plane parameters are nearly equal. The c/a ratio for the three compounds, Ca$_3$Co$_2$O$_6$,
Ca$_3$CoRhO$_6$, and Ca$_3$CoIrO$_6$ are 1.143, 1.166 and 1.189 respectively and it appears that the magnetic behavior critically depends on  relative strengths of interchain and intrachain coupling determined by c/a ratio in some fashion.    

To conclude, we have identified a stoichiometric spin-chain compound, Ca$_3$CoIrO$_6$, 
undergoing a glassy-type of magnetic ordering 
below 30 K. We have compared and contrasted the observed magnetic behavior with those of isostructural compounds, Ca$_3$Co$_2$O$_6$ and Ca$_3$CoRhO$_6$. It is particularly notable that the title compound does not exhibit PDA structural characteristics unlike the other two compounds. It may also be added that the behavior of this compound (glassy magnetic transition with a single magnetic transition temperature, which is also robust to the application of magnetic field) is  unique among this class of quasi-one-dimensional magnetic oxides.  The results sufficiently demonstrate that this class of compounds 
provide an avenue to probe the consequences of geometric frustration in 
psuedo-one-dimensional magnetic compounds, presumably resulting from the relative strengths of interchain versus intrachain magnetic coupling. 

The authors would like to thank K.K. Iyer for his help while 
performing the experiments.


\end{document}